\documentclass[fleqn,usenatbib,linenumbers]{mnras}

\usepackage{newtxtext,newtxmath}

\DeclareRobustCommand{\VAN}[3]{#2}
\let\VANthebibliography\thebibliography
\def\thebibliography{\DeclareRobustCommand{\VAN}[3]{##3}\VANthebibliography}

\usepackage{graphicx}	
\usepackage{amsmath}	
\usepackage{bm}		
\usepackage{pdflscape}	

\newcommand{\fuv}{{\rm{FUV}}}
\newcommand{\nuv}{{\rm{NUV}}}
\newcommand{\g}{{\rm{G}}}
\newcommand{\bp}{{\rm{BP}}}
\newcommand{\rp}{{\rm{RP}}}
\newcommand{\teff}{{T_{\rm{eff}}}}
\newcommand{\gunit}{{\rm{cm\ s^{-2}}}}
\newcommand{\chired}{{ \chi^2_{\rm{red}} }}
\newcommand{\Lwd}{{L_{\rm{WD}}}}
\newcommand{\lbol}{{L_{\rm{bol}}}}
\usepackage{longtable}
	
\usepackage{color, colortbl}
\usepackage[first=0,last=9]{lcg}

\newcommand{\gaia}{{\em{Gaia}}}
\newcommand{\galex}{{GALEX}}

\title[WD--MS binaries within 100 pc]{Hunting Down White Dwarf--Main Sequence Binaries Using Multi-Wavelength Observations}

\author[P. K. Nayak et al.]{
Prasanta K. Nayak,$^{1,2}$\thanks{E-mail: nayakphy@gmail.com}
Anindya Ganguly,$^{1}$
and Sourav Chatterjee$^{1}$
\\
$^{1}$Tata Institute of Fundamental Research, 
Homi Bhaba Road, Colaba, Old Navy Nagar,  
Mumbai, India 400005\\
$^{2}${Instituto de Astrofísica, Pontificia Universidad Católica de Chile, Av. Vicuña MacKenna 4860, 7820436, Santiago, Chile}
\\
}

\date{Accepted 2023 November 14. Received 2023 October 28; in original form 2022 December 12}

\begin{document}
\label{firstpage}
\pagerange{\pageref{firstpage}--\pageref{lastpage}}
\maketitle

\begin{abstract}

Identification of white dwarfs (WD) with main-sequence (MS) companions and characterization of their properties can put important constraints on our understanding of binary stellar evolution and guide the theoretical predictions for a wide range of interesting transient events relevant for, e.g., Rubin Observatory (LSST), ZTF, and LISA. In this study, we combine ultraviolet (UV) and optical color-magnitude diagrams (CMDs) to identify unresolved WD--MS binaries. In particular, we combine high-precision astrometric and photometric data in the optical from \gaia\ -DR3 and UV data from GALEX GR6/7 to identify 93 WD--MS candidates within 100 pc. Of these, 80 are newly identified. Using the Virtual Observatory SED Analyzer (VOSA) we fit the spectral energy distributions (SEDs) of all our candidates and derive stellar parameters, such as effective temperature, bolometric luminosity, and radius for both companions. We find that our identification method helps identify hotter and smaller WD companions (majority with $\geq$10,000 K and $\leq$0.02 $R_\odot$) relative to the WDs identified by past surveys. We infer that these WDs are relatively more massive (median $\sim$ $0.76\,M_\odot$). 
We find that most of the MS companions in our binaries are of the $K$ and $M$ spectral types.

\end{abstract}

\begin{keywords}
 (stars:) binaries: general ; (stars:) Hertzsprung-Russell and color-magnitude diagrams ; (stars:) white dwarfs; (Galaxy:) solar neighborhood; virtual observatory tools 
\end{keywords}



\section{Introduction} \label{sec:intro}
The most common dark companion to the main sequence (MS) star in a binary system is a white dwarf (WD). However, identifying them is challenging even in the Solar neighborhood as WDs are typically low-luminosity objects. 
Identifying WD--MS binaries in the Solar neighborhood is important to not only understand binary stellar evolution but also for the completeness of the observational sample. A large and homogeneous sample of WD--MS binaries is also useful to guide binary population synthesis studies to put meaningful constraints on several uncertain aspects of binary stellar evolution \citep[][]{Toonen2017,torres_2022} which may have implications for a variety of interesting stellar populations including Type Ia supernovae \citep[][]{Wang2012}, and post-common envelope binaries \citep[PCEBs;][]{Toonen2013, Camacho2014, Zorotovic2014, cojocaru_2017}.   
Identifying a large sample of WD--MS binaries can help constrain theoretical models that can predict rates of transient events that would be detectable in large numbers in the LSST, ZTF, and LISA surveys. 

Once identified, followup studies with photometric variation \citep[e.g.,][]{Shakura_1987,Masuda2019}, or radial velocity \citep[RV, e.g.,][]{Zeldovich_1966, Trimble_1969, Chawla2022, WDS1_parsons2016, WDS2_Rebassa2017, WDS5_Ren2020, hernandez2022} , can help constrain the orbital parameters of WD--MS binaries. This can in turn help put constraints on different formation channels for these binaries. For example, in the case of WD--MS binaries in wide orbits, both companions evolve independently without affecting each other's individual evolution and the more massive companion evolves to become a WD \citep[][]{Garcia1997, Farihi2010_WD-RD}. Studying such wide systems can be used to constrain the age-metallicity relation in the solar neighborhood \citep[][]{Rebassa2016b, Rebassa2021}, the secondary-mass function \citep[][]{Ferrario2012}{}{} or the age-activity-rotation relation \citep[][]{morgan2012,Rebassa2013a, Skinner2017}{}{}.  
On the other hand, if the WD--MS progenitor binary's orbit is sufficiently compact, mass transfer due to Roche lobe overflow (RLOF) plays a crucial role in the stellar and orbital evolution and the binary can go through a common envelope (CE) evolution. Studying the compact WD--MS binary populations is also important in constraining the efficiency of common envelope evolution and its energy budget \citep[][]{Zorotovic2010, Rebassa2012b, Camacho2014, Zorotovic2014}, the critical mass ratio for stable vs unstable mass transfer \citep[e.g.][]{Bobrick_2017}{}{},  
and the mass-radius relation of white dwarfs \citep[][]{Parsons2017}. Furthermore, the WD--MS binary properties bear signatures of its evolution. Mass gain by a WD can push it beyond the limit and create SN~Ia, a very important transient source in astronomy. On the other hand, episodic or continuous mass loss from the WD's progenitor through CE or RLOF, may create the so-called extremely low mass WDs (ELM-WDs) with masses $\le 0.3\,M_\odot$ \citep{istrate_2014_a,istrate_2014_b, nandez_2015}.  
Wind mass accretion may also be important in close-orbit systems which can create WDs with unusual properties \citep[][]{Balman2020}. 
Hence, understanding the demographics of WD binaries in general, and WD--MS binaries in particular, is interesting to ultimately understand the formation of a large range of stellar exotica and constrain the details of binary stellar evolution. 

There have been many studies dedicated to identifying a large number of WD--MS binaries. One way to identify the spatially unresolved systems is through radial velocity measurements through spectroscopic observations. For example, Sloan Digital Sky Survey \citep[SDSS;][]{York2000_sdss, Eisenstein2011_sdssIII, Rebassa_2010_wd_sdssVII, Rebassa2012a, Rebassa2013b, Rebassa2016a} and the Large Sky Area Multi-Object Fiber Spectroscopic Telescope Survey \citep[LAMOST;][]{Cui2012_LAMOST, Chen2012_LEGUE_LAMOST, Ren2014_wdms_LAMOST_DR1, Ren2018_wdms_LAMOST_DR5} have detected the most extensive catalog of over a few thousand spectroscopic WD--MS binaries with M-dwarf companions. Although, a majority of these WDs are in wide systems, a large number (a few hundred) of likely PCEBs too have been reported.   
Combining  spectroscopic observations from LAMOST or Radial Velocity Experiment \citep[RAVE;][]{Kordopatis2013_RAVE_DR4, Kunder2017_RAVE_DR5} with ultraviolet photometry \citep[from GALEX, Galaxy Evolution Explorer;][]{martin2005_GALEX} a large number of WD--MS binary candidates were identified with MS companions spanning a large range in spectral types from A to K \citep[][]{WDS1_parsons2016, WDS2_Rebassa2017, Anguiano2020_wd_APOGEE, WDS5_Ren2020}. 
The Optical-UV combined survey used $T_{\rm{eff}}$--UV-color (FUV$-$NUV) relation to separate WD--MS candidates from MS stars. However, any spectroscopic survey is limited by its rather stringent limiting magnitude. On the other hand, recently, astrometric solutions in the Tycho-\gaia\ astrometric solution \citep[TGAS;][]{Michalik2015_TGAS}, \gaia-DR2 \citep[][]{Evans2018_gaia_DR2_photometry}, and \gaia-DR3 \citep[][]{Gaia_DR3_NSS} have been used to identify WD--MS candidates. Although, these candidates need further UV or spectroscopic observations for confirmation of their candidature. 

Though the number of identified unresolved WD--MS binaries has increased from a few tens to over a few thousand in the last three decades, 
the sample is still incomplete even within a local volume of 100 pc \citep[][; hereafter RM21]{Rebassa2021}. For example, most recently, RM21 identified 112 WD--MS binaries within 100 pc located in the gap region between the MS and WD sequence on the optical color-magnitude diagram (CMD). Although, their study did not include MS--WD binaries hidden in the MS region, their analysis showed that $\approx91\%$ of WD--MS binaries may actually be hidden within the MS. Our aim in this paper is to identify the WD--MS binaries hidden within the MS of the optical CMD in a volume-limited sample below a fixed distance $100$ pc to improve the completeness \citep{Inight_2021, Jimenez-Esteban_2018_GaiaDR2_WD, Jimenez-Esteban_2023_GaiaDR3_WD}.

In a WD--MS binary, the WD is expected to have its peak emission in the UV, while the emission from the MS star should peak in the optical. If there are significant differences in the optical and UV fluxes from the WD and the MS, an unresolved WD--MS binary would appear on the MS region in the optical CMD but pop up on the WD sequence in the UV CMD. We apply this to identify potential candidate WD--MS binaries using their locations on optical and UV CMDs. In order to know the accurate position of an object on the CMD, one must know the distance and extinction. 
We use optical data from \gaia, which provides G, BP, and RP magnitudes as well as accurate astrometry and parallax which help estimate the distances to the stars with less than a few percent in error \citep[][]{Gaia_DR1_2016, Gaia_DR2_2018, Gaia_eDR3_2021, Gaia_DR3_catalog_validation}. For UV, we use GALEX GR6+7 data \citep[][]{bianchi2017}. Note that, active MS stars, while on the MS on the optical CMD, may also show-up in the FUV-bright region on a UV CMD. Nevertheless, it is expected that the majority of the WD--MS binaries appear hotter (FUV--NUV $<$ 3.5 mags) compared to the active MS stars \citep[][]{Anguiano_2022}, thus limiting contamination.

In \autoref{sec:methods}, we discuss the data sets used in this study, the details of the method we employ to identify the WD--MS binary candidates, and the tools to estimate their stellar parameters. 
We discuss our results, a separate method to estimate WD mass, and a comparison with previous literature in \autoref{discussion}. We summarize our results and conclude in \autoref{conclude}.


\begin{figure*}
	
	\includegraphics[width=2\columnwidth]{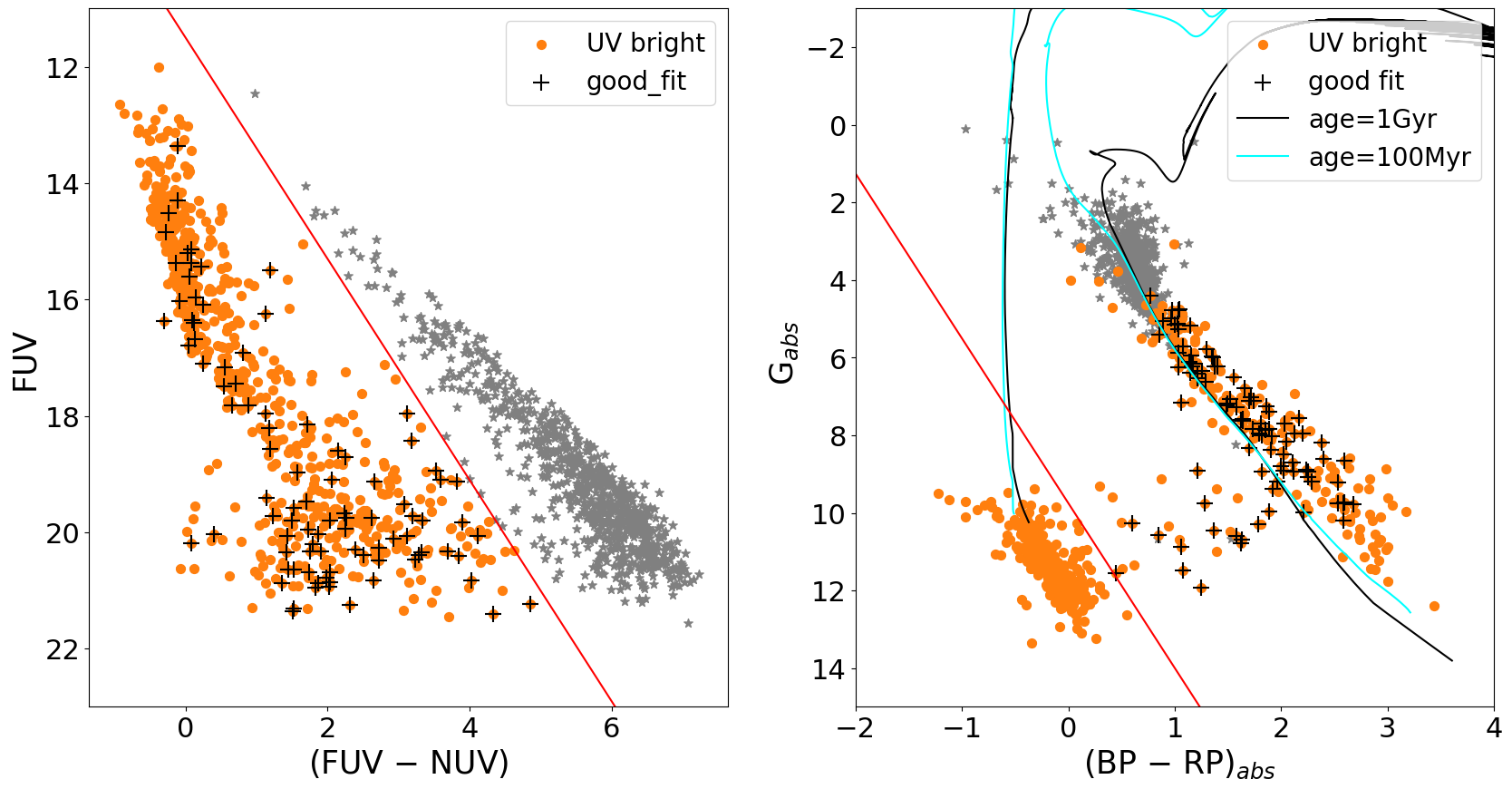}
    \caption{\textit{Left:} UV (FUV$-$NUV vs FUV) CMD for sources within 100\ pc with both GALEX and \gaia-DR3 observations. The red line separates the hotter and FUV-bright sources (orange points) from the others (grey asterisks) on the UV CMD. \textit{Right}: Optical CMD (BP$-$RP vs absolute G) in the absolute plane, after correcting for distance modulus and extinction for the same sources with the same color and point type. The counterpart of the red line on UV CMD is also shown.  Isochrones of 100 Myr (cyan line) and 1 Gyr (black line) are overlaid to indicate the evolutionary sequences. Grey asterisks populate the MS and post-MS regions on both CMDs. The UV and optical CMDs indicate that the FUV-bright sources (orange points) are mainly WDs. However, a fraction of them migrates to the gap and MS regions on the optical CMD. These are candidate WD--MS binaries. Black plus highlights the candidate WD--MS binaries for which we could completely explain the SED from UV to IR using WD--MS composite model fluxes.  
     }
    \label{selection}
\end{figure*}


\section{Data Selection and Characterization}
\label{sec:methods}
In this section, we detail how we use \gaia\ and \galex\ data to select WD--MS candidate binaries in this study. Furthermore, we discuss how we estimate the stellar properties of both companions of these binaries. 

\subsection{Data Selection} 
\label{data}

\gaia's data release 3 (\gaia\ -DR3) brings a major advancement compared to \gaia\ -DR2 in the systematic errors and astrometric solutions. Precision in parallax is increased by 30 percent, in proper motions it is improved by a factor of 2. The systematic errors in the astrometry were suppressed by 30-40\% for parallax and by a factor of $\sim$2.5 for proper motions \citep{Gaia_eDR3_2021}. 
Therefore, we use \gaia\ -DR3 to first identify the sources located within 100 pc. In addition, we also implemented some photometric and astrometric quality cuts based on the errors associated to parallax ($\varpi$), fluxes in G, BP and RP bands. We consider the sources which follow the following conditions:
\begin{itemize}
\item $\varpi/\sigma_{\varpi} \ge$ 10 
\item $I_{\rm BP}/\sigma_{I_{\rm BP}} \ge$ 10
\item $I_{\rm RP}/\sigma_{I_{\rm RP}} \ge$ 10
\item $I_{\rm G}/\sigma_{I_{\rm G}} \ge$ 10
\end{itemize}

\noindent where, $\varpi$  is the parallax in  arcseconds, $I_{\rm G}$, $I_{\rm BP}$ and  $I_{\rm RP}$ are the fluxes in  the bandpass filters $G$, ${BP}$ and ${RP}$, respectively, and $\sigma$ are the standard errors of the corresponding parameters. 
\gaia\ provides the sky positions of sources in the J2016 epoch, while the other catalogs used in this study use the J2000 epoch. So, we convert \gaia's source coordinates into the J2000 epoch from J2016 using DR3 astrometric solutions in order to crossmatch the catalog with GALEX GR6+7 \citep{bianchi2017} and other optical/IR survey catalogs. We find 15,878 sources common in the \galex\ GR6+7 and \gaia\ -DR3 within 100 pc, which has both FUV and NUV detections. To consider only the sources with reliable photometry, we remove sources with $>0.2$ mags error in both FUV and NUV magnitudes. This condition is equivalent to selecting only detections with $SNR>5$ \footnote{https://www.eso.org/~ohainaut/ccd/sn.html}. We also discard sources flagged as artifacts or extended sources in the \galex\ catalog. Applying these two source selection criteria, we are left with 2,634 sources. 
\footnote{Cut-offs in the renormalized unit weight error (RUWE) and astrometric excess noise (AEN) are often used in the literature for source selection. We intentionally avoid using these parameters since it has been shown that binarity may lead to high values of RUWE and AEN \citep[][]{Belokurov2020}. } 
Then we crossmatch our sample with the following surveys: APASS DR9 \citep{apass} and PanSTARRS-DR2 \citep{magnier_panstarrs_dr2} for optical, and 2MASS \citep{2mass}, Spitzer, and ALLWISE \citep{allwise} for IR observations. We only consider sources with counterparts in all of the above surveys. Out of the 2,634 sources, we find 1,463 sources of interest satisfying all of the above conditions.

\subsection{Identification of Candidates} 
\label{sec:identification}
We now analyze these 1,463 sources to identify WD--MS binary candidates using UV and optical CMDs. 
The left panel of \autoref{selection} shows the UV CMD of the 1,463 sources of interest. The orange dots indicate the FUV-bright hotter population compared to those marked with grey asterisks. We separate this hotter population from the others using the red line given by $\fuv = 1.9\times(\fuv-\nuv) + 11.5$. The right panel shows the same sources in the optical CMD using absolute magnitudes after taking into account the distance modulus obtained from \gaia's -DR3 catalog \citep[][]{bailer-jones_2021edr3} and extinctions obtained from a three dimensional (3D) Dust Map Based on \gaia\ -DR2, Pan-STARRS 1, and 2MASS \citep[MWDUST;][]{Drimmel_2003_mwdust, Marshall_2006_mwdust, Bovy_2016_mwdust, Green_2019_mwdust}. We first estimate the 3D reddening (E(B$-$V)) using the MWDUST python package\footnote{https://github.com/jobovy/mwdust} based on the distance and Galactic positions of the sources of interest and then determine the extinction (A$_V$) considering extinction curve with R$_V= $ 3.1. Comparing the UV and optical CMDs, we can clearly see that the majority of the hot FUV-bright population identified in the UV CMD (orange points) remains where the WDs are expected to be found on the optical CMD, on the left of the red line. However, a fraction of the FUV-bright sources show up in the MS region and in the gap between the MS and the WD cooling sequence. The red line separating WDs from the rest population is defined as $\g_{abs} = 4.25\times(\bp -\rp)_{abs} + 9.75$. 
This clearly indicates that these sources are candidate unresolved WD--MS binaries. When the WD dominates the contribution in the UV and the MS dominates the contribution in the optical, then the source appears among the hot WDs in the UV CMD but shows up among the MS stars in the optical CMD. On the other hand, if the WD also contributes somewhat in the optical, the source can also appear in the gap region between the MS and the WD cooling sequence in the optical CMD. 
We find 257 candidates from this shift in CMDs within our sources of interest.  
\autoref{selection} also shows two isochrones for ages 100\,Myr and 1\,Gyr, generated using the \textbf{MIST} isochrone package \citep{mist0, mist1} on the optical CMD to demonstrate the MS and WD evolutionary sequences as reference.


\begin{figure*}
	\includegraphics[width=1.0\columnwidth]{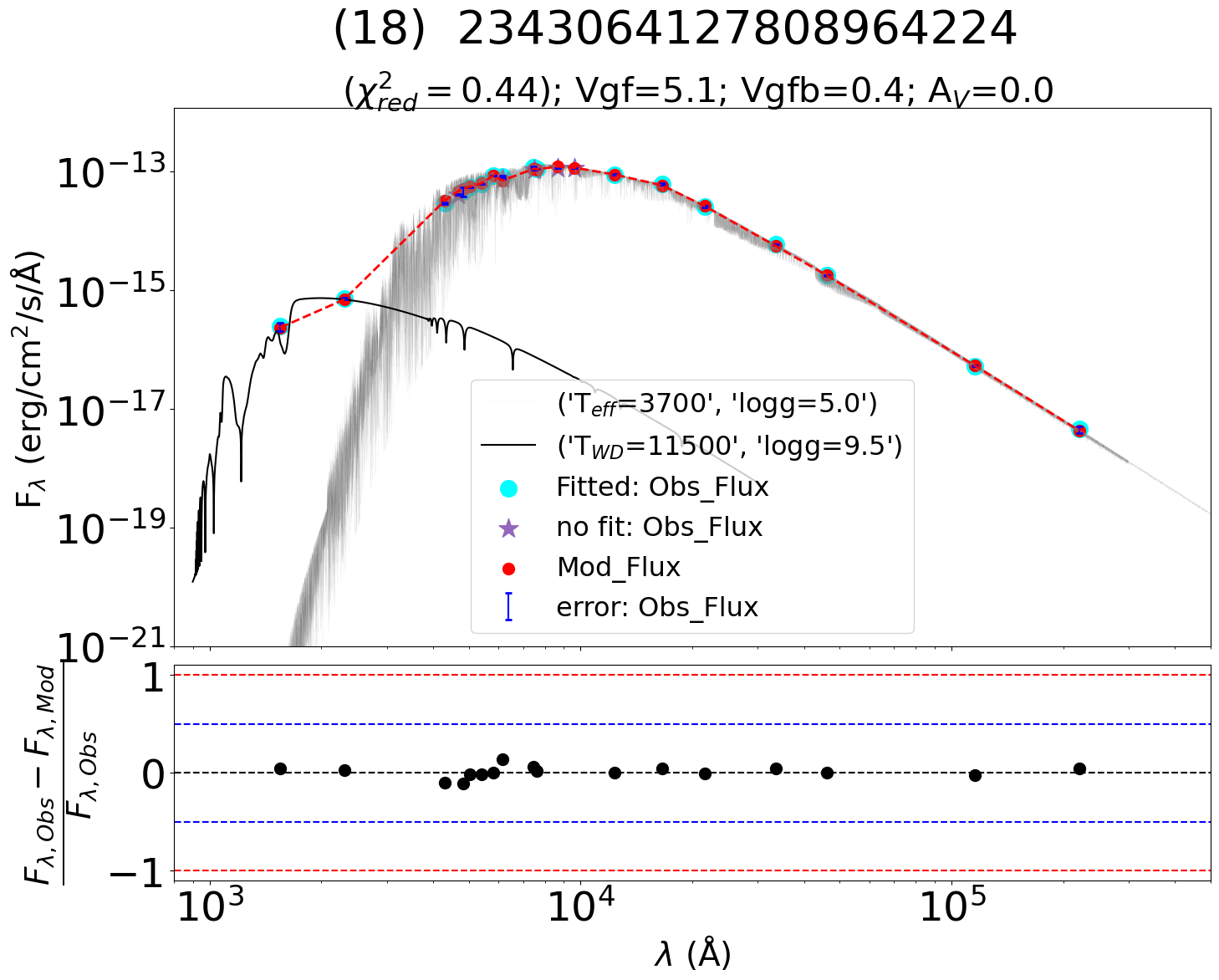} \includegraphics[width=1.0\columnwidth]{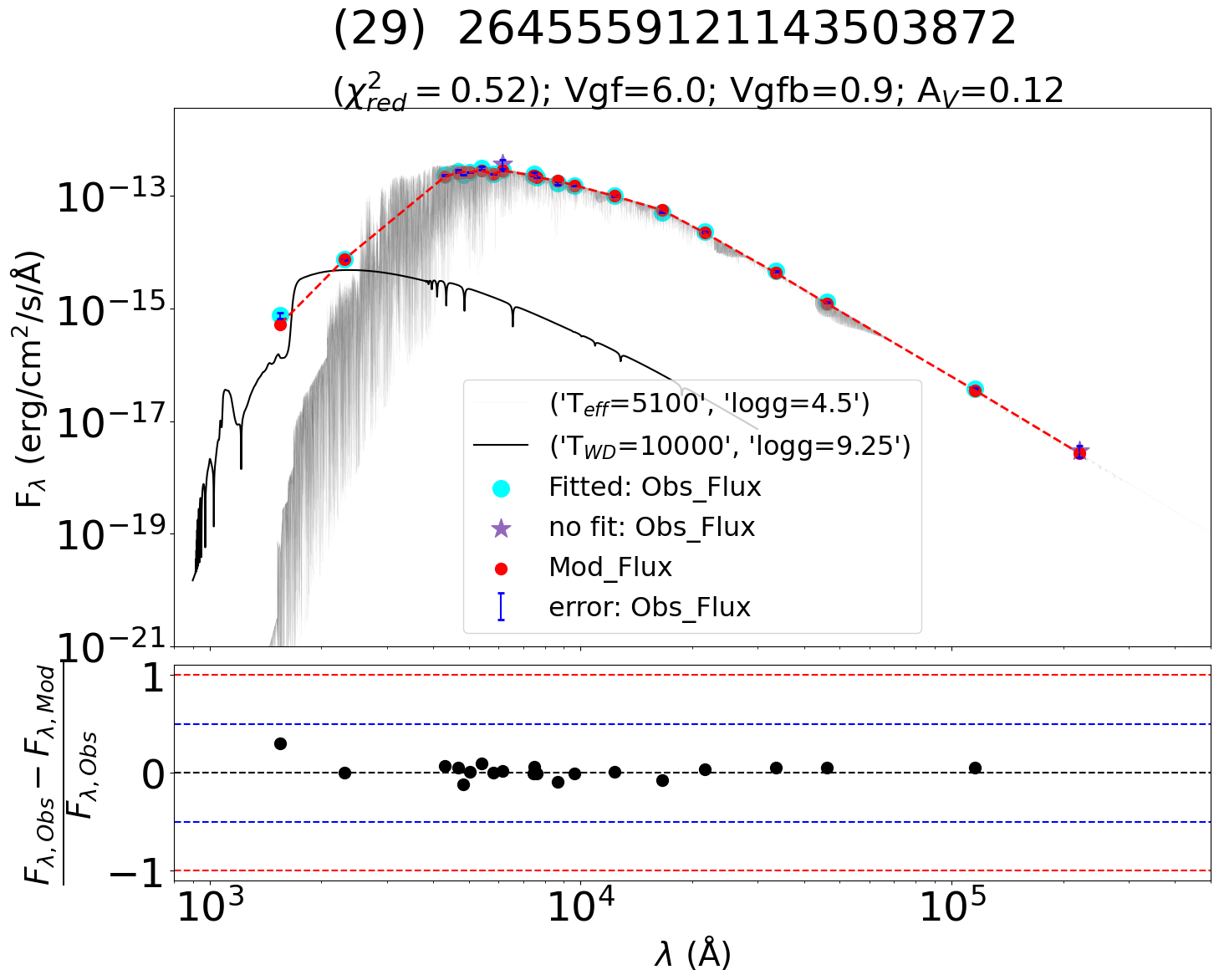} 
    \caption{Two examples of well-fitted SEDs of WD--MS binaries. Index number as per \autoref{tab:source_table}, \gaia\ -DR3 source ID, the values of A$_V$, $\chired$, Vgf and Vgf$_b$ are mentioned on top of each panel. (Top panel:) Cyan points (with blue errors) denote the observed flux from UV to IR. The observed data points with no errors or large errors ($>$0.2 mag) are marked as asterisks and are not included in the fit. The black (grey) line represents the best-fit synthetic spectra of WD (MS). 
    The red points indicate the expected combined model fluxes from the best-fit synthetic spectra. (Bottom panel:) Fractional residue fluxes are shown in different bands. The blue and red dashed lines represent 50\% and 100\% residue flux, while the black line represents zero residue. 
    There is excellent agreement between the cyan and red points, and less than 50\% residue fluxes in every bands indicating a well-fitted SED. 
    }
    \label{sed}
\end{figure*}

\begin{figure*}
	\includegraphics[width=1.0\columnwidth]{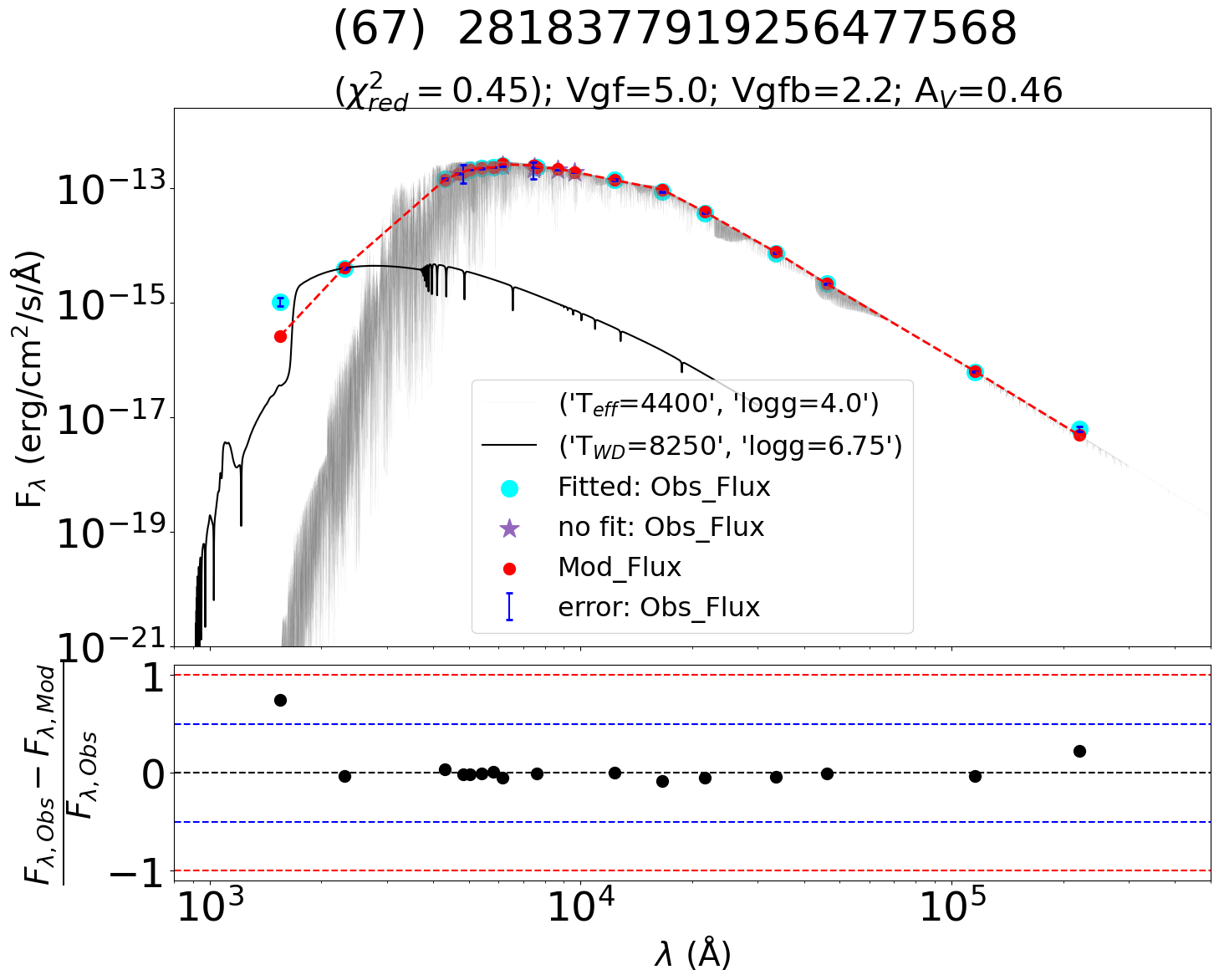} \includegraphics[width=1.0\columnwidth]{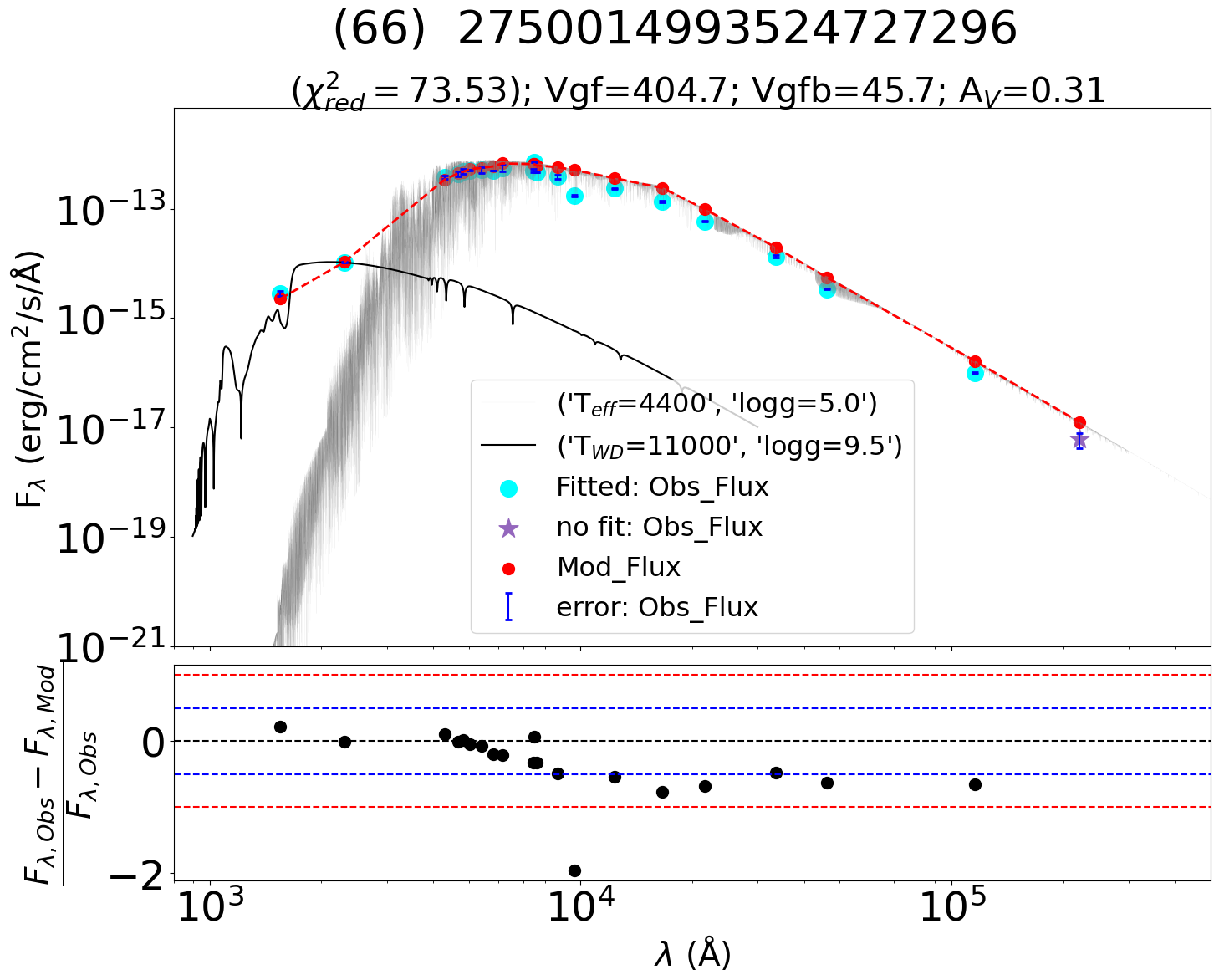} 
    \caption{Same as \autoref{sed} but for two representative examples of badly-fitted SEDs of WD--MS candidate binaries. Index number as per the online catalog, \gaia\ -DR3 source ID, the values of A$_V$, $\chired$, and Vgf$_b$ are mentioned at the top. In the case of source 67 (left), the model spectra are unable to fit the observed FUV flux. In contrast, for source 66 (right), the model spectra are able to fit the fluxes in the UV region but not the optical or IR regions. 
}
    \label{badfit}
\end{figure*}

\subsection{Characterization of WD--MS Binaries} 
\label{sec:wd_params}
We now proceed to test these 257 candidates to ascertain whether they indeed are WD--MS binaries by analyzing their spectral energy distributions (SEDs). 
We use the virtual observatory SED analyzer's \citep[VOSA;][]{bayo2008} binary fit algorithm to fit the observed SEDs with theoretical SEDs created using WD and MS star models to extract the best-fit stellar parameters for both companions simultaneously. We use the BT-Settl-CIFIST models \citep{Baraffe2015} for the MS star. These models provide a large database of spectra spanning effective temperature $1,200\leq\teff/\rm{K}\leq7,000$ with a resolution of 100\,K, $0\leq\log\ g\leq5.5$ with a resolution of 0.5.\footnote{Throughout the paper, $g$ is in $\gunit$.} For our purposes, we use the full range for $\teff$ available in VOSA, but restrict ourselves to $4\leq \log\ g\leq5$ since MS stars are expected to have $\log\ g$ within this range. We use WD evolution models of \citet{Koester2010} included in the VOSA toolkit which provides spectra for H-rich WDs in the range $5,000\leq \teff/{\rm K}\leq80,000$ with varying resolutions between 250 and 10,000\,K, $6.5\leq\log\ g\leq9.5$ with a resolution of 0.25. 

As input parameters to the VOSA toolkit, we provide names, coordinates, distances, A$_V$, and magnitudes in different filters for our sources. 
Since our sources are all within 100\,pc by selection, we consider solar metallicity for all of them. 
We do not include photometric data points with unreported errors or errors $>0.2$ mag during our binary SED fitting. 
VOSA performs multiple iterations by varying $\teff$, $\log\ g$, and scaling factor ($M_d$) to minimize $\chi^2$ to find the best-fit spectra to the observed flux distribution. 
The scaling factor $M_d\equiv (R/D)^2$ is used to scale the model flux to match the observed flux, where $R$ and $D$ denote the radius and distance of the source. In our exercise, since the distance is already known from \gaia, the scaling factor gives us the radius.

After performing WD--MS binary fit using VOSA, we find that the SED analysis provides well-fitted SEDs for 94 of the 257 candidates. For these 94 candidates, we are able to constrain the properties of both components. We are not able to use reduced $\chi^2$ ($\chired$) as a parameter to determine the quality of the fit, since, in several cases, though visual inspection suggests them to be well-fitted SEDs, their $\chired$ values are large. This is a known problem in VOSA and arises when the photometric data points have very small observational flux errors (say $<$1\% of observed flux). So, even if the model reproduces the observation apparently well, the deviation can be much higher than the reported observational error (increasing the value of $\chired$ ). To mitigate this, VOSA has introduced the parameter called the visual goodness of fit (Vgf$_b$) by modifying the $\chired$ formula, where the error is considered to be at least 10\% of its observed flux\footnote{http://svo2.cab.inta-csic.es/theory/vosa/helpw4.php?otype=star\&action=help}.  
The parameter Vgf$_b$ provided by VOSA with a value $\leq$15 is usually considered as a proxy for well-fitted SEDs (RM21). 
However, from visual inspection, we notice that even with Vgf$_b$ $<$ 15 some SEDs were not fitted to our satisfaction.
Therefore, we impose an additional criterion on absolute fractional residual flux $f_{\rm{residue}}$ = $\lvert (F_{\lambda,\rm{obs}}-F_{\lambda,\rm{model}})/(F_{\lambda,\rm{obs}}) \rvert$, where $F_\lambda$ is the observed flux at waveband $\lambda$ and subscripts `obs' (`model') denotes the observed (best-fit model) flux. 
We impose that in the UV region, both FUV and NUV data must individually satisfy $f_{residue}$ $<0.5$. In addition, in the optical-IR region, more than 80\% of data points must have the $ f_{residue} $ $<0.5$. For example, if 14 data points are used in the optical-IR region for the SED fitting, the model should fit at least 12 data points with $ f_{residue} $ $<0.5$. This small difference in the criteria for UV and optical-IR regions is due to a statistically large number ($\geq$14) of data points used for SED fitting in the optical-IR region compared to the UV region with just two data points.     
So, we consider the SEDs to be well-fitted when the sources satisfy both the Vgf$_b$ and the $ f_{residue} $ criteria simultaneously. 
We provide the \gaia\ -DR3 and GALEX IDs, and stellar parameters of these 94 sources in the online catalog. We have also listed $\chired$, Vgf$_b$, and $ f_{residue} $ in FUV \& NUV bands in the Table for reference. A representative part of the catalog is presented in \autoref{tab:source_table}. 

We show the SEDs of two well-fitted WD--MS binaries in \autoref{sed} as examples. 
The Cyan points with blue error bars indicate the observed fluxes in different filters. The observed data points with no errors or large errors ($>$0.2 mag) are marked as asterisks and are not included in the fit. The black and grey spectra represent the best-fit WD and MS models to the observed SEDs, respectively. The combined model fluxes in corresponding filters are marked as red points and connected with the red dashed line. 
Bottom panel of each plot shows $f_{residue}$ in different bands. The match between the cyan and red points, and $ f_{residue} $ of less than 50\% in all the bands indicate that these are good fits according to our criteria. 
At the top of each plot, we mention the index number as per \autoref{tab:source_table}, \gaia\ -DR3 source ID, $\chired$, Vgf, Vgf$_b$, and A$_V$ values for each candidate. 
Similar figures for all the well-fitted WD--MS binary candidates are available online.

The SEDs of the other 163 candidates could not be fitted well with WD--MS composite model spectra using VOSA according to our criteria. SEDs of the 163 candidates and a catalog containing their positions, \gaia\ -DR3 and GALEX IDS are made available online, and two sample SEDs are shown in \autoref{badfit}. The color codes used in this figure are the same as mentioned in the \autoref{sed}, except the numbers within parentheses above each plot follow the index number from the online catalog for which WD--MS models fit badly to the observed SEDs. The left panel shows an example where although the Vgf$_b<15$ and $\chired<1$, there is a large excess flux in FUV. In contrast, the right panel of \autoref{badfit} shows an example where the best-fit model spectra fits the UV region of the SED well but not the optical and IR regions. 
The reason for why these candidates could not be fitted well by the WD--MS binary model can be varied, including but not limited to improper cross-match, observed UV excess coming from stellar activity of the MS star and not from a WD companion, and contamination from nearby stars in the optical/IR. This example also illustrates why simply using $\chired$ or Vgf$_b$ as calculated by VOSA as the only parameter for the goodness of fit can be problematic in an analysis similar to this.


\begin{figure}
	\includegraphics[width=\columnwidth]{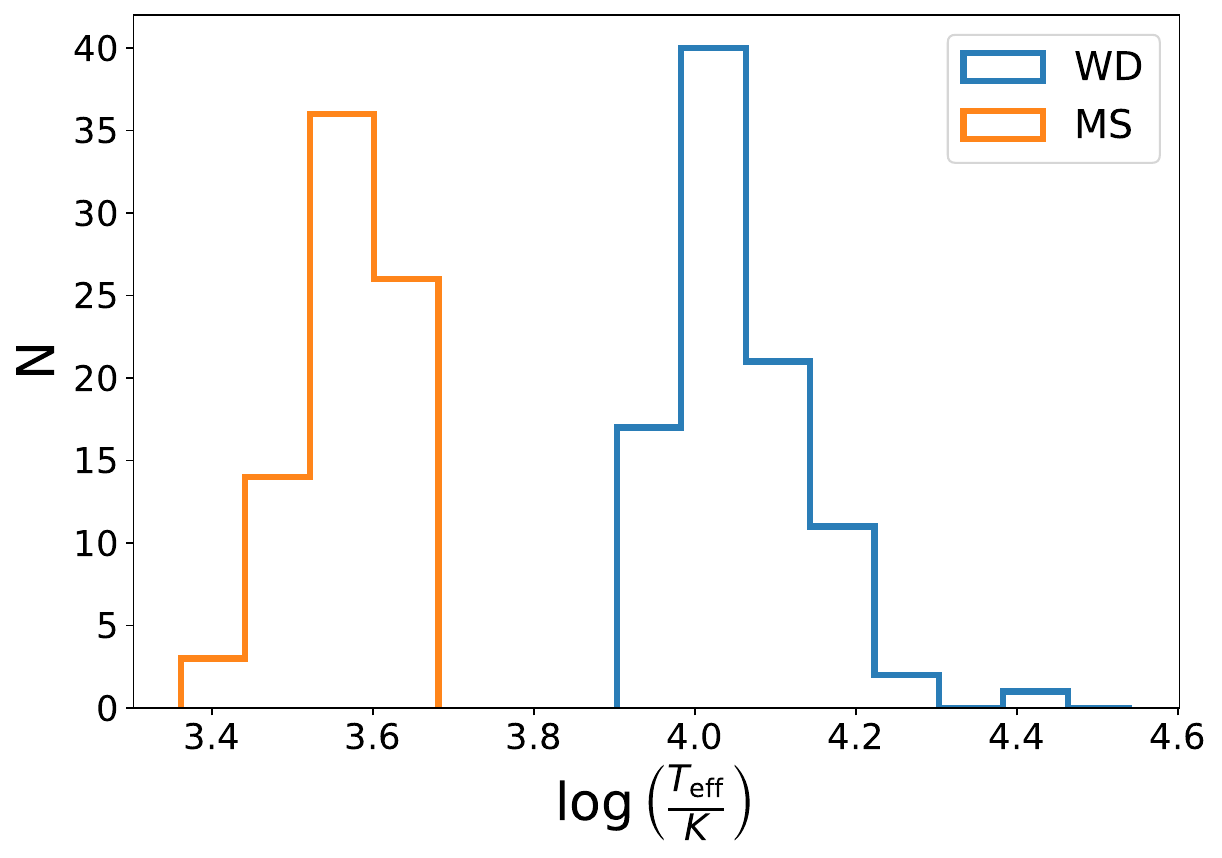}
    \caption{Distribution of $\teff$ for the WDs (blue) and MS (orange) stars in our candidate WD--MS binaries. The median, 25th, and 75th percentiles for the MS s (WD) is $\teff/\rm{K}=3800^{+775}_{-300}$ ($11000^{+1750}_{-1000}$). 
    }
    \label{teff}
\end{figure}

\begin{figure}
	\includegraphics[width=\columnwidth]{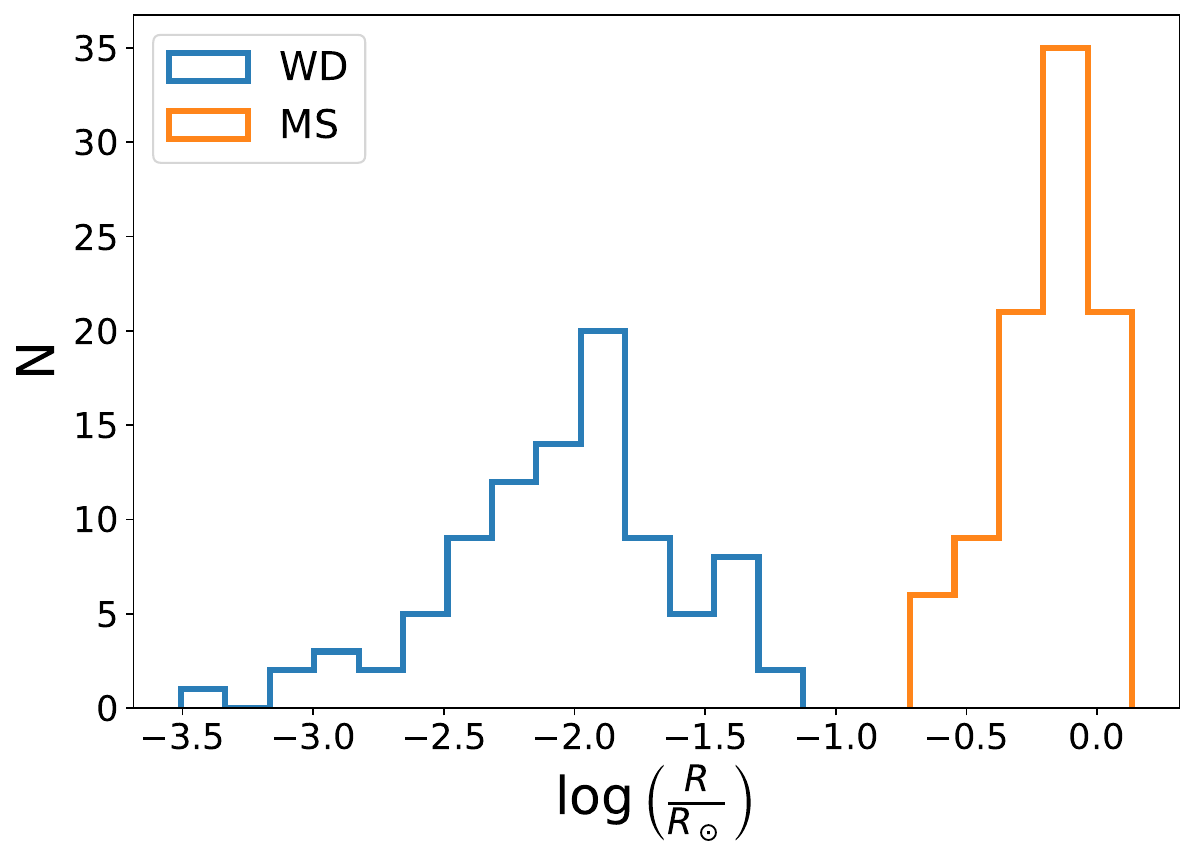}
    \caption{Same as \autoref{teff}, but showing the stellar radii for WDs (blue) and MS stars (orange). The median, 25th, and 75th percentiles for the MS stars (WDs) is $\log(R/R_\odot)=-0.16^{+0.10}_{-0.14}$ ($-2.02^{+0.23}_{-0.27}$). 
    }
    \label{radius}
\end{figure}


\section{Results and Discussion} \label{discussion}
The 94 sources for which we could fit the observed SEDs using model spectra from WD--MS binaries are marked with black plus signs in \autoref{selection}. We find that while 80 of these sources lie on the MS in the optical CMD, the others populate the gap region between the MS and the WD cooling sequence. 
WD--MS binary fits using the VOSA tool provide us with best-fit stellar parameters for both companions. 
In particular, $\teff$ and radius estimates from SEDs are robust \citep{bayo2008}. All stellar properties of each source are summarized in \autoref{tab:source_table}. We show the distributions of $\teff$ and radii for the MS stars (orange) and the WDs (blue) in \autoref{teff} and \autoref{radius}. We find that in our sample, the MS (WD) stars have $\teff/\rm{K}=3800^{+775}_{-300}$ ($11000^{+1750}_{-1000}$). The corresponding numbers for the radii of the MS (WD) stars are $\log(R/R_\odot)=-0.16^{+0.10}_{-0.14}$ ($-2.02^{+0.23}_{-0.27}$).\footnote{The numbers in all cases denote the median. The lower and upper errors denote the 25th and 75th percentiles, respectively.} Thus, most of the MS components in our catalog of WD--MS binaries are K and M spectral types based on the $\teff$ vs spectral type relation \citep{pecaut2013}.

\subsection{WD mass}
\label{S:wd_mass}
As mentioned on the VOSA website, the fitting process and the predicted flux are relatively less sensitive to $\log\ g$. This poses a challenge to put direct constraints on the masses of the components since a slight change in $\log\ g$ can create a large difference in the estimated mass. Hence, instead of directly using the $\log\ g$ values given by VOSA, we estimate the masses of the WD companions using the WD evolutionary models created by \citet{Bedard2020} assuming that the WDs in our sample have CO cores with Hydrogen atmosphere. Using $\teff$ and bolometric luminosity ($\Lwd$) of the WDs from our SED fits and WD evolutionary models we estimate the masses and cooling ages.

\begin{figure}
	\includegraphics[width=1.0\columnwidth]{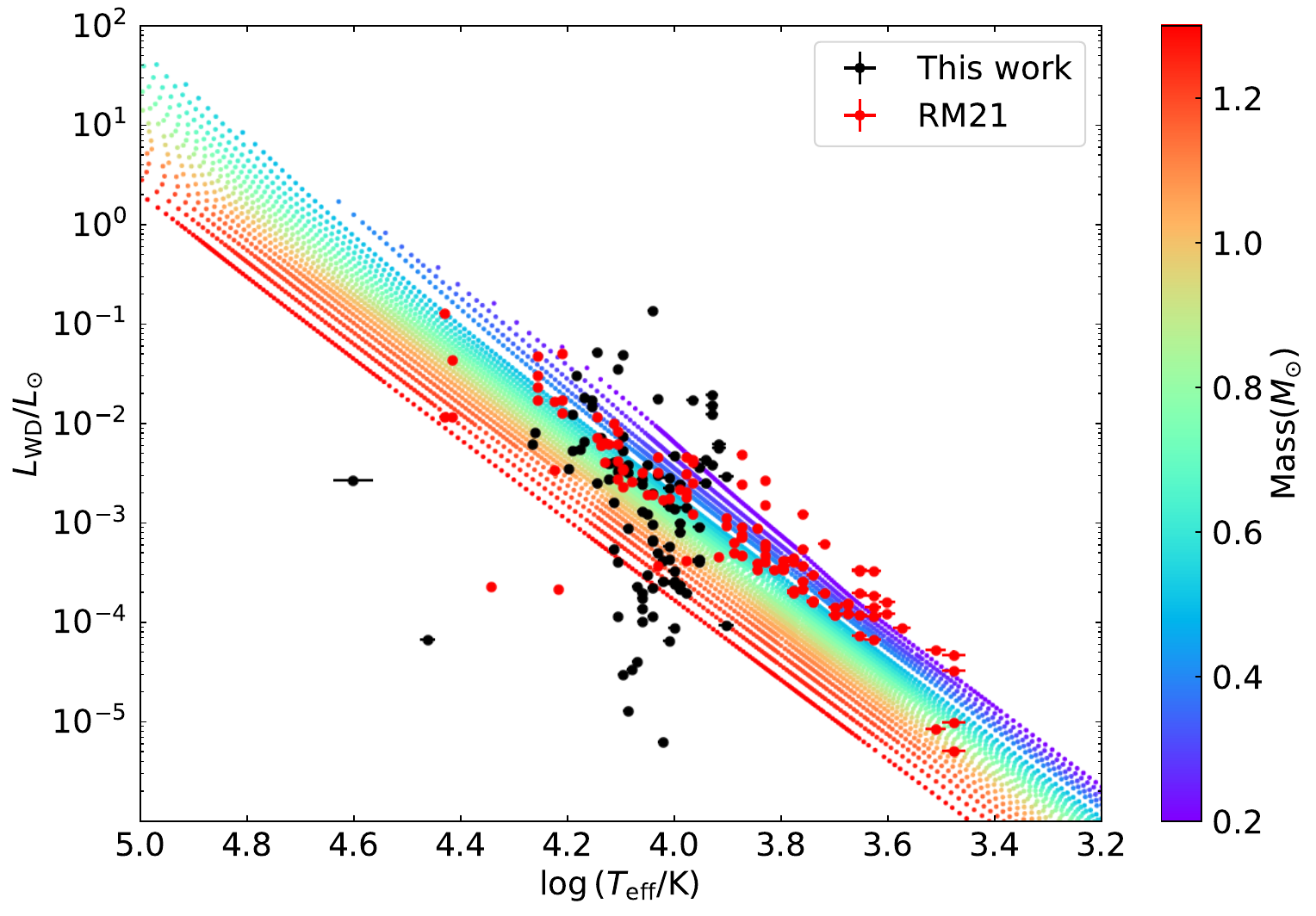} 
    \caption{$\teff$ vs $\Lwd$. Model cooling sequences for CO WDs with a Hydrogen atmosphere are shown where the color code denotes WD mass \citet{Bedard2020}. The black and red dots are the WDs identified in this work and in RM21, respectively.   
    }
    \label{wd_mass}
\end{figure}

In \autoref{wd_mass}, we show $\Lwd$ as a function of $\teff$ for the WDs in our sample with black dots. WD cooling tracks from models \citep[][]{Bedard2020} are overlaid. The cooling tracks are color-coded with WD mass. In general, for the same cooling age, lower-mass WDs would have lower $\teff$. On the other hand, over time, the WDs would cool along the sequences shown in the figure for any particular mass. 
First, we draw 2D boxes around the $\Lwd$--$\teff$ estimates from VOSA for each of these sources. The width and height of these boxes come from the errors associated with the estimated $\teff$ and $\Lwd$. We estimate the masses of the WDs from the model cooling tracks using linear interpolation adopting the central points of these boxes. Errors in the estimated WD masses are obtained using the extreme values of these 2D boxes.
We find that 14 WDs in our sample have $\Lwd$ above the model evolutionary tracks. These are expected to have lower masses compared to the limiting mass $0.2\,M_\odot$ in CO-core WD models provided by \citet{Bedard2020}. As a result, we could not estimate the masses of these high-$\Lwd$ WD candidates in our sample.  
We also find 18 of our candidate WDs with $\Lwd$ below the model evolutionary tracks. We refrain from estimating their masses also because we do not want to extrapolate.

We find that the masses of the WDs (where we could estimate) are distributed in the range $\sim$0.2 to 1.3 $M_\odot$ (\autoref{wd_mass}) and $\log\ g$ values are between 6.5 to 9.5. The individual masses of these candidate WDs are listed in \autoref{tab:source_table}. Interestingly, we find 5 WD candidates with masses $\le0.3\, M_\odot$. Such low-mass WDs are not expected to form via isolated single stellar evolution. Instead, it is expected that these low-mass WDs form through mass transfer from the WD's progenitor during its RGB phase via RLOF or via CE evolution, known as ELM-WDs \citep[][]{istrate_2014_a,istrate_2014_b,nandez_2015,li_2019, Brown2022, ambreesh2022}. 
Previous spectroscopic surveys such as the SDSS magnitude-limited sample \citep[][]{Rebassa_2010_wd_sdssVII, Rebassa2012a_wd_sdssXIV, Rebassa2013a} contain only a few such low-mass WDs with MS stars as their companion. These are hard to detect due to their low-mass companions and low luminosity. About $31\%$ of the more recent catalog from RM21 contains potential ELM-WDs with MS companions. Though many ELM-WD+MS candidates have been recently detected \citep[][]{Jadhav_2019, Jadhav_2023,  Subramaniam_2020, Khurana_2023}{}{}, 
their formation channels need better understanding and we encourage detailed population studies in this regard.

\begin{figure}
	\includegraphics[width=1.0\columnwidth]{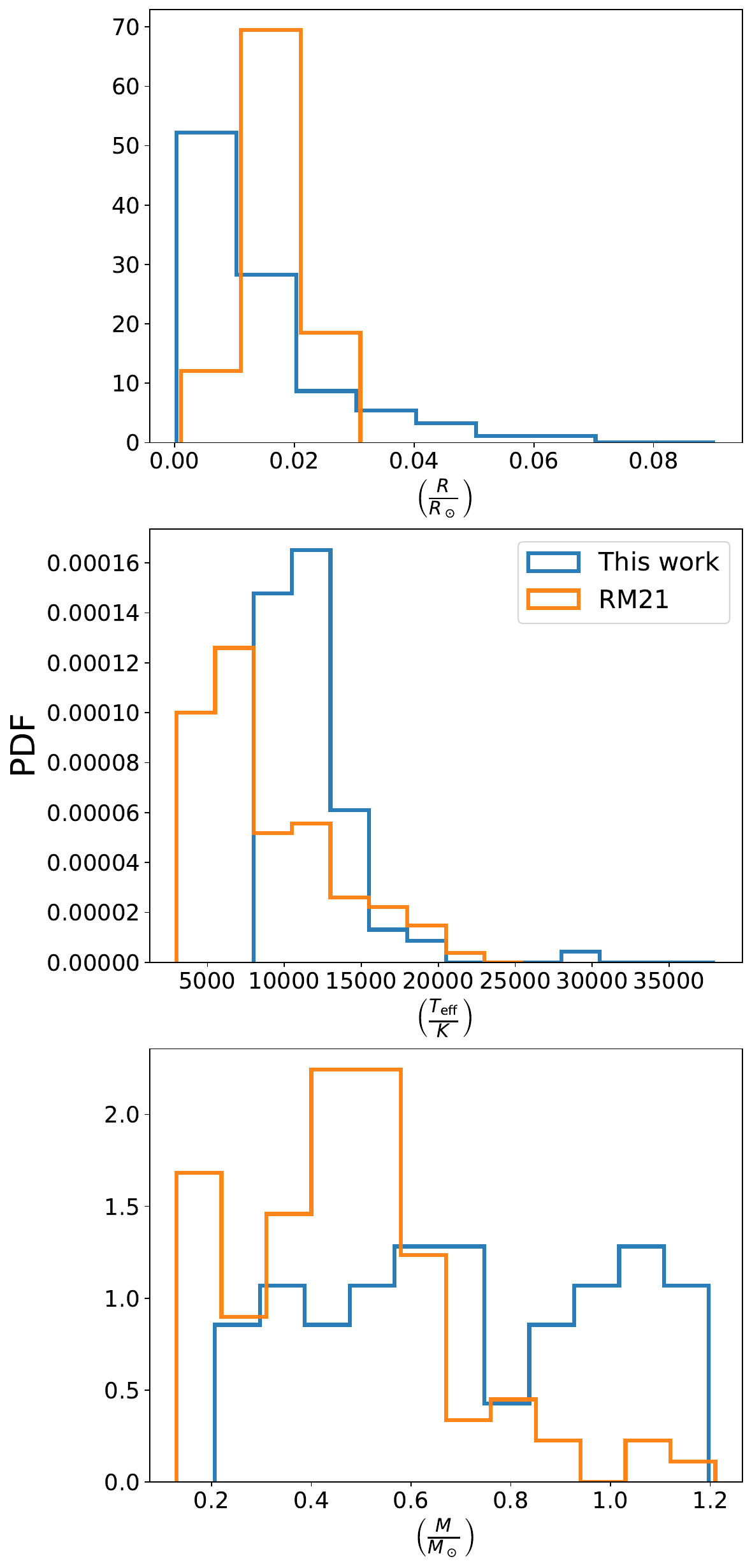}
    \caption{ Comparison of properties for WDs identified in this work (blue) and RM21 (orange). From top to bottom, we show distributions of radius, $\teff$, and mass, respectively. }
    \label{compare}
\end{figure}


\subsection{Comparison with existing catalogs} \label{comparison}
The most recent WD--MS catalog provided by RM21 targeted candidates within 100 pc using \gaia\ -DR2, similar to the distance cut-off we have used in this study. However, their targets were sources that appear between the MS and the WD cooling sequence on the optical CMD of \gaia. As a result, they were more sensitive to sources where both the WD and the MS have shared contribution towards the total optical flux, and all of their 112 WD--MS candidates reside in the gap between MS and WDs on \gaia 's CMD. In contrast, our targets were selected via a shift between the UV CMD, where our candidates reside completely where the WDs are expected to, and the optical CMD, where the majority (80 out of a total 94) of our WD--MS candidates reside on the MS, while the others reside in the gap region. As a result, our catalog is complementary to that of RM21.  
\autoref{compare} compares the estimated mass, radius, and $\teff$ of the WDs identified in this study (blue) with those found in RM21 (orange). It is clear that we detect hotter, heavier, and smaller WDs in our catalog of WD--MS binaries compared to those found by RM21. 

Perhaps, the WDs identified in the RM21 catalog can be compared to the subset of those we find in the gap region between the MS and the WD cooling sequence on \gaia's optical CMD. The RM21 catalog contains 112 WD--MS binaries located in the gap region. We find fewer sources (14) in the gap region. We investigate the reason by implementing our selection criteria on the RM21 catalog. We find that only 34 sources out of the their 112 have both FUV and NUV magnitudes in GALEX GR6+7, a primary criteria for our selection (\autoref{sec:identification}). After removing stars with FUV and NUV artifacts and flags, and with errors in FUV and NUV magnitudes $>$ 0.2 mags, we are left with 23 sources. Then we crossmatch these 23 sources with 2MASS, ALLWISE, APASS, and PanSTARRS surveys. We find that only 16 sources have counterparts in all these surveys. After applying the selection criteria of moving locations between the UV and optical CMDs, we get 13 sources that follow our selection criteria for WD--MS binary candidates. The remaining 3 remain in the WD region both in the UV and the optical CMDs, hence are not considered as candidates in our catalog. 
Thus, the relatively fewer candidates (14) in the gap between WDs and MS on the optical CMD in our catalog can be attributed to our additional data selection criteria relative to RM21. While these 13 common sources are part of our catalog, we find well-fitted WD--MS SEDs only for 7 sources. Among the remaining 6 sources from the RM21 catalog, we find 5 sources have Vgf$_b$ $<$ 15, but $f_{\rm{residue}}>50\%$ and one source has Vgf$_b$ $>$ 15.
Overall, our catalog contains 14 WD--MS binaries located in the gap region with well-fitted SEDs. Thus, in effect, we find 7 new WD--MS binaries in the gap region not present in the RM21 catalog. 

We further notice that our estimated MS properties agree well with those estimated by RM21 for these 7 common sources. However, there are some differences in the WD properties; our estimated $\teff$ is somewhat higher compared to those found by RM21. We find that RM21 did not correct for A$_V$ in their SED fits. Since the UV region is more affected by small changes in A$_V$ (A$_{FUV}$ = 2.7 $\times$ A$_V$ \& A$_{NUV}$ = 2.8 $\times$ A$_V$), our reddening corrected estimates are expected to be closer to reality compared to the estimates of RM21.

Another recent and large WD--MS binary survey is the White Dwarf Binary Pathways Survey (WDBPS) which combined astrometric solutions from TGAS and \gaia\ -DR2, and radial velocity solutions from the LAMOST and RAVE surveys to identify 814 WD--MS binaries \citep[][]{ WDS5_Ren2020}. When we crossmatch the catalog with TGAS, \gaia\ -DR3, and \galex, we find only 786 matches with both FUV and NUV observations. Out of 786, only 47 are located within 100 pc based on \gaia\ -DR3 parallax. If we remove sources with potentially bad observations in GALEX based on the FUV \& NUV flags in the catalog and those with FUV and NUV errors $>0.2$ mag, we are left with 26 sources. This indicates that $\approx$45\% of all sources in the WDBPS catalog by \cite{ WDS5_Ren2020} with distance $<$ 100 pc are associated to larger uncertainties. 
Nevertheless, we find 15 sources after crossmatching the remaining 47 with 2MASS, ALLWISE, APASS, and PanSTARRS catalogs. Our UV and optical CMD relative locations-based criteria are satisfied by 9 of these 15. Of these 9 sources, we could find well-fitted SEDs for only 3. 
Overall, in our catalog of 94 WD--MS binaries, 80 are located on the MS in the optical CMD. Thus, our catalogue increases the number of WD+MS binaries located in the MS locus of the optical CMD for distances under 100 pc.

There have been several other previous attempts to identify WD--MS binaries in the Solar neighborhood using various photometric, astrometric, and spectroscopic observations, as described in the \autoref{sec:intro}. We also cross-match our WD--MS catalog with those from the literature. 
We find that 5 WD--MS candidates in our catalog match with the catalog of WD--MS binaries detected from LAMOST \citep{Ren2018_wdms_LAMOST_DR5}. 
There is one WD--MS binary from our catalog already identified by the SDSS spectroscopic catalog of WD--MS binaries \citep[][]{Rebassa2012a}. We also find a few common sources between our catalog and previous catalogs of WD--MS binaries detected via various photometric surveys. For example, there is one common source between our catalog and that of \citet{Rebassa2013b} which used color selection criteria by combining optical data from SDSS-DR8 (ugriz), infrared data from UKIRT Infrared Sky Survey (yjhk), 2MASS (JHK) and/or WISE (W1W2) to identify 3,419 photometric WD--MS candidates harboring cool white dwarfs with dominant M-dwarf companions.  
Comparison with \gaia\ -DR3 non-single star (NSS) catalog \citep[][]{Gaia_DR3_NSS} provides five match with ours. 
The match with NSS catalog also highlights the importance of multi-wavelength study from UV to IR to constrain the binary properties for sources in the NSS catalog \citep{Ganguly_2023}.

Active MS stars may populate the low-temperature range of the WD region in the UV CMD and can also move from the FUV bright region to the MS in the optical CMD. However, they are also expected to have X-ray counterparts. We checked with the active MS catalog \citep[][]{Boro_Saikia_2018_active_MS, Martinez_2010_active_MS}{}{} and for their X-ray counterparts in Chandra \citep[][]{Evans_2010_CHANDRA}{}{}, XMM-Newton \citep[][]{Saxton_2008_XMM-Newton-slew, Webb_2020_XMM-Newton-epic}{}{}, and ROSAT all-sky survey data \citep[][]{Boller_2016_2RXS} to find only one match. We discard this source from our catalog and mark it with a star. We also did not include this star in Figures \ref{teff}, \ref{radius}, \ref{wd_mass}, and \ref{compare}. 
These common candidates between our catalog and those in the literature provide confirmation of their candidature, as well as validate our method for identifying new WD--MS candidates. Overall, we find that our catalog contains 93 WD--MS binary candidates and out of them, 80 are newly identified. 

Based on the simulations by RM21, $\sim$91\% of the total WD--MS population in our Galaxy is expected to be hidden in the MS of the optical CMD since the MS companion dominates the total optical flux. Our study shows an easy-to-implement pathway to reveal this population given that multi-wavelength data exist for these sources. 
Since the key ingredient of our detection method is to construct both the UV and the optical CMDs using the same sources, our results show that the census of WD--MS binaries, particularly those harboring hotter, younger, more massive, and smaller WDs can be improved in future through deep UV photometric observations \citep[][]{WDS1_parsons2016, WDS2_Rebassa2017, WDS5_Ren2020}. Considering the challenges to carry out high-resolution spectroscopic observations of a large number of stars in the field, our method illustrates a simple way to detect WD--MS binaries and estimate their stellar parameters. Of course, once identified, parameter estimation can be significantly improved by follow-up high-resolution spectroscopic studies \citep[][]{WDS1_parsons2016}. 
In this context, we expect that deep UV surveys with the Ultra-Violet Imaging Telescope on board AstroSat \citep[][]{tandon2017a, tandon2017b} or the upcoming UV mission INdian Spectroscopic and Imaging Space Telescope (INSIST) \citep[e.g. ][]{purni_2022_INSIST} can provide a huge boost to identifying new WD--MS binaries in the Milky Way by simply adopting the method described in this study.

\section{Summary and Conclusions}
\label{conclude}
We have used UV and optical CMDs and the relative locations of specific sources on these two CMDs as a tool to identify unresolved WD--MS binaries. Combining high-precision astrometric and photometric data from \gaia\ -DR3 with GALEX GR6/7 we have identified 93 WD--MS candidates. Out of these 93 candidates, 80 are new. 

We have used a binary SED fitting algorithm within VOSA to estimate the stellar parameters such as $\teff$, luminosities, and radii of both binary components. We have found that our method preferentially identifies binaries with relatively hotter (median $\sim11,000\,\rm{K}$) and smaller WD companions (median $\sim 0.01\,R_\odot$)  
with respect to previous catalogs \citep[e.g.,][]{Rebassa2021}, thus creates a complementary set. 

We have used the WD evolutionary models of \citet{Bedard2020} for CO core and Hydrogen atmosphere to estimate the mass and $\log\ g$ for the WD components. 
WD masses in our catalog are between $\sim$0.2 and 1.3 M$_\odot$ and $\log\ g$ spans between 6.5 and 9.5. This study brings out a population of ELM-WDs ($\le 0.3\,M_\odot$), which are likely created due to mass loss from their progenitor stars. Although many ELM-WD binaries have been detected previously, these sources typically have normal WD companions (e.g., \citealt{Brown2022}, but for an overall collection of known ELM-WD binaries, see \citealt{Khurana_2023, Jadhav_2019, Jadhav_2023, Subramaniam_2020}). ELM-WDs with MS companions such as those in our catalog are less explored in the literature and we encourage further theoretical and observational studies.

In conclusion, our study describes an easy pathway to identify a large number of WD--MS candidates by combining \gaia\ data with UV photometry. We believe that with the existing \citep[UVIT;][]{tandon2017a,subra2016} and upcoming \citep[INSIST;][]{purni_2022_INSIST} UV missions coverage of deep UV photometry will improve. The method adopted in this study will significantly improve the completeness of the WD--MS binary census in the solar neighborhood. 

\section*{Acknowledgements}
PKN acknowledges TIFR's postdoctoral fellowship. PKN also acknowledges support from the Centro de Astrofisica y Tecnologias Afines (CATA) fellowship via grant Agencia Nacional de Investigacion y Desarrollo (ANID), BASAL FB210003. AG acknowledges support from TIFR’s graduate fellowship. SC acknowledges support from the Department of Atomic Energy, Government of India, under project no. 12-R\&D-TFR-5.02-0200 and RTI 4002. All simulations were done using TIFR HPC. 

\section*{Data Availability}
All data will be freely available in online mode\footnote{https://zenodo.org/records/10149072}.


\vspace{5mm}
\section*{Software}
\texttt{Astropy}\ \citep{astropy:2013, astropy:2018}; \texttt{matplotlib}\ \citep{matplotlib}; \texttt{numpy}\ \citep{numpy}; \texttt{scipy}\ \citep{scipy}; \texttt{pandas}\ \citep{pandas}; \texttt{VOSA}\ \citep{bayo2008}; \texttt{mwdust}\ \citep{Green_2019_mwdust}.



\bibliographystyle{mnras}
\bibliography{WDMS} 


\begin{landscape}
 \begin{table}
  \caption{WD--MS binaries and component properties. 
  A truncated list of MS--WD candidates. The full list is available online. We provide \gaia\ and \galex\ IDs, source coordinates, and MS and WD properties.}
  \tabcolsep2.90pt $ $
  \label{tab:landscape}
  \begin{tabular}{c|c|c|cc|ccc|cccc|c|c|cc|}
    \hline
  & \gaia\ ID & \galex\ ID & RA & DEC &  &   MS Properties&  &  & & WD Properties& & $\chired$ & Vgf$_b$ &  & $ f_{residue} $ \\
  \cline{6-12}
  & &  & (deg) &
 (deg) & $\teff$ & Radius & $\lbol$ &$\teff$& 
 Radius& $\lbol$& Mass & & &  &  \\
  &  &  &  &
  & ($10 ^{3}$K) & $(R_{\odot})$ & $(10^{-2} \times L_{\odot})$ & ($10^{3}$K) & 
 $(10^{-3}R_{\odot})$ &$(10^{-3}L_{\odot})$ & $(M_{\odot})$ &  &  & FUV & NUV \\
    \hline
1 &  1015888309580557056 &   J085900.8+493519 & 134.7535557 & 49.58853147 & 3.6 $\pm$0.05 & 0.512 $\pm$0.002 & 4.06 $\pm$0.04 & 15.75 $\pm$0.12 & 7.78 $\pm$0.03 & 3.38 $\pm$0.05 & 1.03 $\pm$0.01 & 1.84 & 0.8 & 0.02 & 0.03 \\
2 &  101612981989146752 &   J022326.7+224405 & 35.8615023 & 22.73471064 & 3.9 $\pm$0.05 & 0.617 $\pm$0.0 & 7.82 $\pm$0.03 & 11.25 $\pm$0.12 & 4.43 $\pm$0.0 & 0.29 $\pm$0.0 & 1.28 $\pm$0.01 & 11.34 & 3.7 & 0.44 & 0.11 \\
3 &  1058663641228841600 &   J104234.7+644205 & 160.6450061 & 64.70157966 & 3.6 $\pm$0.05 & 0.567 $\pm$0.001 & 4.92 $\pm$0.03 & 13.0 $\pm$0.12 & 7.58 $\pm$0.01 & 1.54 $\pm$0.02 & 1.03 $\pm$0.02 & 1.29 & 1.1 & 0.05 & 0.13 \\
4 &  1117595952649491968 &   J091254.9+692152 & 138.2291746 & 69.36423524 & 4.5 $\pm$0.05 & 0.904 $\pm$0.001 & 30.27 $\pm$0.28 & 10.25 $\pm$0.12 & 7.53 $\pm$0.01 & 0.57 $\pm$0.01 & 1.04 $\pm$0.03 & 0.21 & 1.5 & 0.28 & 0.01 \\
5 &  1119989933060503680 &   J093653.1+714553 & 144.2209351 & 71.76483602 & 4.8 $\pm$0.05 & 0.653 $\pm$0.0 & 20.35 $\pm$0.19 & 9.5 $\pm$0.12 & 5.14 $\pm$0.0 & 0.19 $\pm$0.0 & 1.24 $\pm$0.02 & 0.37 & 0.7 & 0.31 & 0.01 \\
\vdots &  \vdots & \vdots &\vdots & \vdots & \vdots & \vdots & \vdots & \vdots & \vdots & \vdots & \vdots & \vdots & \vdots & \vdots & \vdots \\
    \hline
  \end{tabular}
\label{tab:source_table}
 \end{table}
\end{landscape}



\bsp	
\label{lastpage}
\end{document}